\documentclass[aps,prl,twocolumn,groupedaddress, showpacs]{revtex4-1}

\newcommand{\bra}[1]{\left\langle{#1}\right\vert}
\newcommand{\ket}[1]{\left\vert{#1}\right\rangle}
\usepackage{graphicx}

\begin{document}


\title{Multi-Colour Quantum Metrology with Entangled Photons}


\author{Bryn Bell$^1$}


\author{Srikanth Kannan$^1$}
\author{Alex McMillan$^1$}
\author{Alex S. Clark$^2$}
\author{William J. Wadsworth$^3$}
\author{John G. Rarity$^{1,}$}
\email[]{john.rarity@bristol.ac.uk}
\affiliation{$^1$Centre for Quantum Photonics, Department of Electrical and Electronic Engineering, University of Bristol,  Bristol BS8 1UB, UK}
\affiliation{$^2$Centre for Ultrahigh bandwidth Devices for Optical Systems (CUDOS), Institute of Photonics and Optical Science, University of Sydney, NSW 2006, Australia}
\affiliation{$^3$Centre for Photonics and Photonic Materials, Department of Physics, University of Bath, Bath BA2 7AY, UK}


\date{\today}

\begin{abstract}
Entangled photons can be used to make measurements with an accuracy beyond that possible with classical light. While most implementations of quantum metrology have used states made up of a single colour of photons, we show that entangled states of two colours can show supersensitivity to optical phase and path-length by using a photonic crystal fibre source of photon pairs inside an interferometer. This setup is relatively simple and robust to experimental imperfections. We demonstrate sensitivity beyond the standard quantum limit and show super-resolved interference fringes using entangled states of two, four, and six photons.
\end{abstract}

\pacs{42.50.Dv, 03.67.Bg}

\maketitle

The measurement of phase in an interferometer is a consequence of the wave-like nature of light, while the uncertainty in that measurement comes from the particle nature of light, revealed in the detection process. This single photon nature of interference was first noted by Taylor's experiment of 1909 \cite{Taylor} while Paul Dirac, in his book on quantum mechanics \cite{Dirac} went as far as to say \textit{`each photon then interferes only with itself'}. Seminal experiments have since shown that this description is too restrictive \cite{PfleegorandMandel}, and that two photon interference is a valid phenomenon \cite{hom}. However multiphoton interferometery experiments aimed at metrology \cite{rarity, highnoon} can shed further light on Dirac's statement. Here, we investigate interferometry with states comprising pairs of photons of distinct wavelengths emitted into one arm or the other of an interferometer. Each photon is detected by a wavelength selective detector and can only have interfered with itself. The correlations between photons ensure that stable interference fringes are only seen when photons are detected in coincidence, with the fringe spacing roughly half their mean wavelength. We extend this super-resolution to four and six photon detections, achieving a fringe spacing one sixth of the original pumping beam wavelength. Subtly, quantum interference between separate photons does play a part in the four and six photon experiments and can affect the shape and contrast of the fringes.
   
\begin{figure}[b]
\vspace{-0.2cm}
\includegraphics[width=\columnwidth]{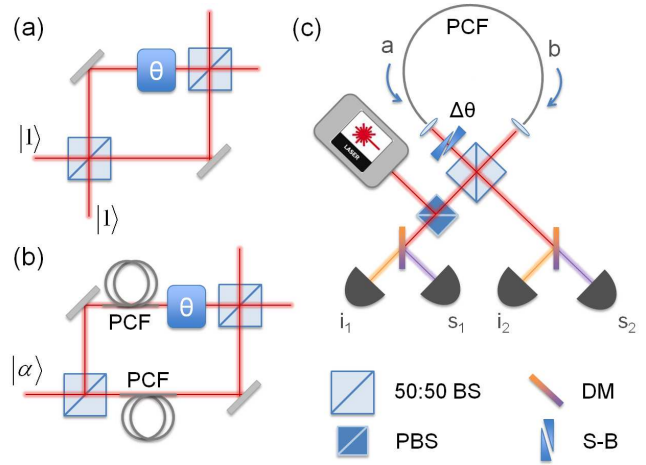}
\vspace{-0.5cm}
\caption{Generation of entangled states for enhanced measurement of $\theta$ (a) by HOM interference between two single photons (b) by a bright coherent state $\alpha$, which pumps two identical PCF pair-photon sources inside the interferometer. The two methods result in equivalent states, except that in (b) the photon pairs do not need to be degenerate in wavelength. (c) Experimental setup using one fibre source of non-degenerate photon pairs pumped in two directions. BS: beam-splitter; PBS: polarizing beam-splitter; DM: dichroic mirror; S-B: Soleil-Babinet compensator.}
\label{setups}
\end{figure}

The uncertainty in measuring an optical phase $\theta$ with an interferometer is limited by Poissonian statistical uncertainty (or shot noise) in the discreet number of photons detected. In classical experiments this standard quantum limit (SQL) on measurement precision is $\Delta\theta\geq 1/\sqrt{N}$, with $N$ the total number of photons detected. In principle, using quantum resources it is possible to dramatically improve on the SQL and reach the Heisenberg limit $\Delta\theta\geq 1/N$ \cite{gioscience}. This has particular applications when the sample has a low damage threshold and it is necessary to extract the maximum possible information without exposing the sample to high intensity illumination.

In our experiment we show potential improvements over the SQL with a setup which overcomes two practical drawbacks common to previous implementations of quantum metrology:

(1) Generating entanglement by Hong-Ou-Mandel (HOM) interference, as in Fig.\ref{setups}(a), is technically demmanding and highly sensitive to any distinguishability between the photons \cite{hom, rarity}. Indistinguishability is usually achieved using narrow filtering, at a cost to the transmission efficiency \cite{Pomarico}. The phase sensitivity of entangled states tends to be highly sensitive to loss, and filtering all photon channels largely cancels out the quantum advantage \cite{Demkowicz}. By generating path entanglement as shown schematically in Fig.\ref{setups}(b), using a source of photon pairs in each interferometer arm, the SQL can be beaten without the need for HOM interference between indistinguishable photons. No quantum resources are required as inputs to the interferometer, only a classical laser.

(2) Using a parametric downconversion source of photons implies that a bright coherent pump beam of half the wavelength is available \cite{nagata, matthews, Xiang}. While entangling two downconverted photons can result in an enhanced precision in terms of a phase, they perform no better than the pump laser at measuring a path-length, as shown below. By using four-wave-mixing (FWM) in photonic crystal fibre (PCF) we produce non-degenerate pairs of photons with the signal and idler equally spaced above and below the pump beam in frequency, so that the central frequency is unchanged and an advantage in length sensitivity is seen using two photon states.

For a given quantum state, the maximum sensitivity to a general parameter $x$ can be calculated from the quantum Cram\'{e}r-Rao bound \cite{gionatphot, luo}, so that assuming a pure state and unitary evolution
\begin{equation}
\Delta x\geq\frac{1}{2\sqrt{\nu} \Delta \hat{H}_x}.
\label{CR}
\end{equation}
Here $\nu$ is the number of copies of the state used and $\Delta\hat{H}_x$ is the spread of an operator $\hat{H}_x$, a Hamiltonian analogue describing the effect of the parameter on the state such that $\ket{\psi}\rightarrow e^{-i\hat{H}_xx}\ket{\psi}$. For a phase $\theta$ applied to a mode, $\hat{H}_{\theta}=\hat{n}$, the photon number operator for that mode. Since we are considering multiple wavelengths, which will experience different phase shifts from a particular sample, a more appropriate choice of parameter is the optical path-length $L$ (an actual length multiplied by a refractive index- we will assume the sample is dispersionless here for simplicity). $\hat{H}_x$ has to be adjusted for this change of parameter and for the presence of multiple modes with frequencies $\omega_k$ and number operators $\hat{n}_k$:
\begin{equation}
\hat{H}_{L}=\sum_k \frac{\omega_k \hat{n}_k }{c}.
\end{equation}
This is proportional to the total energy summed over all wavelengths. Hence we can evaluate the usefulness of a state $\ket{\psi}$ in measuring $L$ from Eq.\ref{CR} by using
\begin{equation}
{\Delta H_{L}}^2 =\bra{\psi}H_{L}^2\ket{\psi} - (\bra{\psi}H_{L}\ket{\psi})^2.
\end{equation}
For $N$ uncorrelated photons of the same frequency, split equally between the sample arm $a$ and a reference arm $b$ of an interferometer,
\begin{equation}
\Delta L\geq\frac{c}{\omega \sqrt{N}},
\end{equation}
showing SQL scaling with a frequency dependence. This expression potentially exaggerates the benefits of using a higher frequency, because as well as showing higher sensitivity, this might be expected to cause more damage to the sample. Rewriting in terms of the total energy detected rather than the number of photons, $E=N \hbar\omega$, we find that for a given energy through the interferometer, higher frequency photons still perform better:
\begin{equation}
\Delta L\geq\left(\frac{c^2\hbar}{\omega E}\right)^{1/2}.
\end{equation}
Using N00N states, maximally correlated states of $m$ photons where $m\ll N$, a $\sqrt{m}$ improvement is possible \cite{highnoon}.
Hence an entangled pair of photons ($m=2$) gives the same information per energy as a coherent state of twice the frequency ($2\omega$ and $m=1$). However, an entangled pair of photons at $\omega\pm\Delta\omega$ can retain a $\sqrt{2}$ advantage over a coherent state with frequency $\omega$.

\begin{figure}[t]
\vspace{-0.4cm}
\includegraphics[width=\columnwidth]{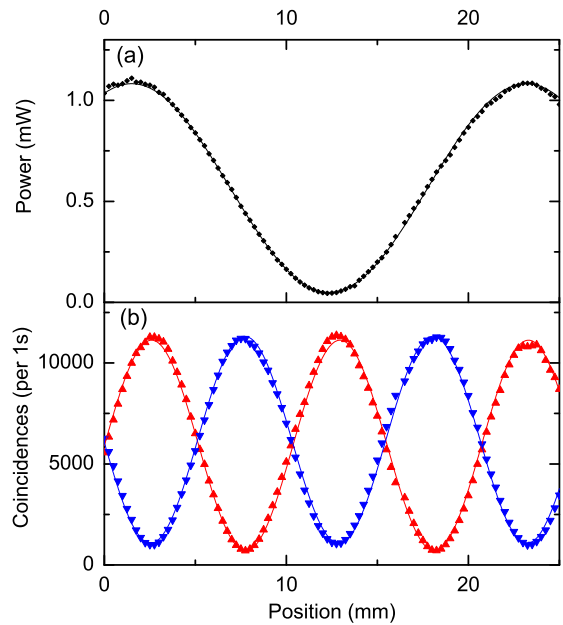}
\vspace{-0.5cm}
\caption{(a) Classical interference with the pump beam at $720~nm$ as the position of the Soleil-Babinet compensator is varied and (b) two photon interference with half the period (red: bunched coincidences; blue: anti-bunched). Both with sinusoidal fit lines.}
\label{fringes1}
\vspace{-0.25cm}
\end{figure}

The proof of principle experiment shown in Fig.\ref{setups}(c) makes use of a Sagnac interferometer to provide intrinsic stability between the clockwise and counter-clockwise paths \cite{li, fulconis, clark, fedrizzi}. Spontaneous FWM in a length of birefringent PCF produces non-degenerate pairs of photons at 625~nm and 860~nm in both paths when pumped in both directions by picosecond laser pulses at 720~nm. Conditional on a single pair being generated, the two photon state can be written as
\begin{equation}
\ket{\psi_2}=\frac{1}{\sqrt{2}}(\ket{10}_{s}\ket{10}_{i}-e^{2i\theta_p}\ket{01}_{s}\ket{01}_{i}).
\end{equation}
The subscripts $s$, $i$, and $p$ denote signal, idler, and pump wavelengths and the kets $\ket{a,b}$ contain the number of photons in interferometer paths $a$ and $b$. Energy conservation in the FWM mixing process requires that $2\omega_p=\omega_s+\omega_i$. The state evolves with a summed phase $2\theta_p=\theta_s+\theta_i=L(\omega_s+\omega_i)/c$. A variable birefringent element allows control of this phase \cite{halder}. The paths are combined at a broadband 50:50 beamsplitter, and at each output the signal and idler modes are separated with dichroic mirrors then detected using silicon avalanche photodiodes.

Classical interference can be observed between the pump light in the two paths (Fig.~\ref{fringes1}(a)). Two photon coincidence detections corresponding to the state $\ket{\psi_2}$ then oscillate as predicted by theory between the bunched and anti-bunched cases (signal and idler emerge from the same or different outputs respectively) in an interference fringe with half the period of the classical case, shown in Fig.~\ref{fringes1}(b). Note that if chromatic dispersion in the sample caused the signal, idler, and pump wavelengths to experience significantly different refractive indices, the period of the two fringes would not be related by an exact factor of two. Despite the relatively large range of wavelengths used, this factor is seen to be two here, in agreement with calculations from the Sellmeier equations for quartz, which predict a deviation of less than $1\%$ \cite{soleilbabinet}.

The visibility of the classical interference is $\sim 92\%$ and mainly limited by different coupling efficiencies through the PCF in the two directions resulting in incomplete cancellation. The two photon visibility is $\sim 88\%$, and is affected to a greater extent by any unmatched coupling efficiencies or losses, as well as background contributions from higher order photon emission. This is well above the usual threshold of $\sqrt{0.5}\approx 70.7\%$ to demonstrate sensitivity better than the SQL \cite{nagata} and corresponds to an uncertainty $\Delta L$ of 0.8 times the SQL. Note that this improvement and the form of the interference fringes are unaffected if spectral correlations are present between signal and idler, leading to them being detected individually in a mixed state. This is an advantage over implementations involving HOM interference to produce entanglement, though in our experiment it is necessary to avoid spectral correlations in order to see any additional improvement from using higher photon number states (see Appendix C).

The two photon detections also suggest that the beamsplitter is acting as a non-unitary operation at the signal wavelength, resulting in some additional phase shifts between interfering terms (see Appendix B). These were taken into account in the theoretical four and six photon curves.


When two signal-idler pairs are created in the PCF, the four photon wavefunction can be written as
\begin{equation}
\ket{\psi_4}=\frac{1}{\sqrt{3}}(\ket{20}_{s}\ket{20}_{i}-e^{2i\theta_p}\ket{11}_{s}\ket{11}_{i}+e^{4i\theta_p}\ket{02}_{s}\ket{02}_{i}).
\end{equation}
Unlike $\ket{\psi_2}$, this behaves differently to a NOON state due to the middle term in the superposition, involving one pair being created in each path - in a NOON state, the photons are either all in one path or the other. $\ket{\psi_4}$ bears more similarity to a Holland-Burnett state \cite{hollandburnett}. In ideal conditions Holland-Burnett states show sensitivity above the SQL but below that of a NOON state, and are an attractive route to entanglement enhanced metrology because they are simple to generate for arbitary $N$ and show a better tolerance to photon loss.
\begin{figure}[b]
\vspace{-1cm}
\includegraphics[width=\columnwidth]{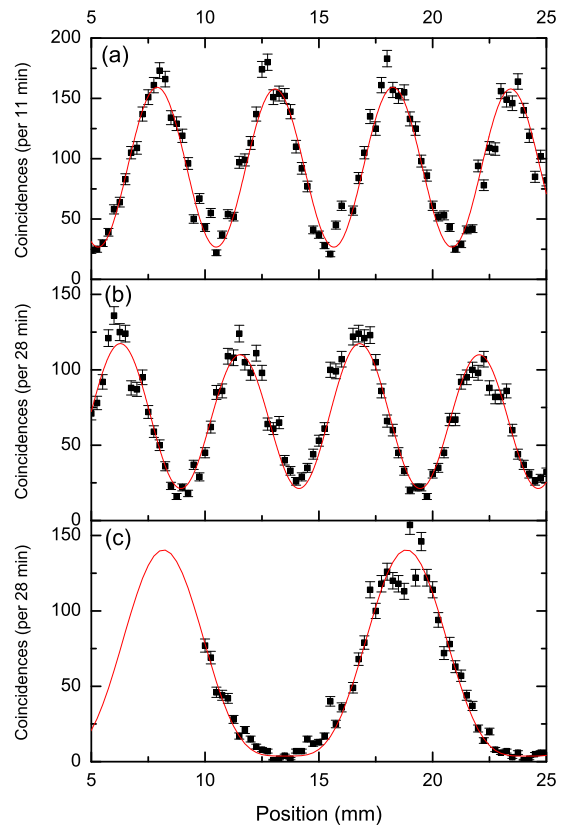}
\vspace{-1cm}
\caption{Four folds counts for (a) the $\ket{11}_s\ket{11}_i$ output (b) the $\ket{20}_s\ket{11}_i$ output and (c) the $\ket{20}_s\ket{02}_i$ output. Fit lines are based on calculation as described in Appendix D.}
\label{fringes3}
\end{figure}

Figure~\ref{fringes3} shows fringes observed when monitoring (at separate times) three different four photon coincidence detections across the outputs of the interferometer. The fringes in Fig~\ref{fringes3} (a) and (b) are approximately sinusoidal with 1/4 the period of the classical fringe, a characteristic of a four photon NOON state. This is because HOM interference acts on the middle term in $\ket{\psi_4}$ and causes the signal photons to bunch together after the final beamsplitter,  so that if the two signal photons are detected at separate outputs, they must have come from one of the other two components of the state, which together resemble a NOON state \cite{nagata}. The same argument applies to detecting the two idler photons at separate outputs, so that any detection pattern involving like photons at separate outputs is expected to show sinusoidal interference with four fold super-resolution. Conversely, for detection patterns with both signals at the same output and both idlers at the same output, the middle term will have an increased effect due to bunching. As seen in Fig.~\ref{fringes3}(c), the effect is a curve with two-fold super-resolution, with some flattening at the minima and sharpening at the maxima due to the presence of a four fold component. This curve is more tolerant to loss, as can be seen from its high visibility compared to (a) and (b), which is expected for interference using a Holland-Burnett like state. Note that the $\ket{20}_s\ket{11}_i$ fringe in Fig.~\ref{fringes3}(b) corresponds directly to the $\ket{31}$ output in the one colour case \cite{nagata}, and similarly it can still show four fold super-resolution with full visibility if separate photon pairs are distinguishable and act independently, in which case the state would be described by $\ket{\psi_2}\otimes\ket{\psi_2'}$ rather than $\ket{\psi_4}$. However, the $\ket{11}_s\ket{11}_i$ fringe in Fig.~\ref{fringes3}(a) does not have an analogue in one colour experiments, and if multi-photon interference did not take place between signals from separate pairs (or idlers from separate pairs) the visibility would be limited to $33\%$. Hence this fringe acts as a test of multiphoton interference taking place and demonstrates that the four photon state can show improved sensitivity compared to multiple two photon states.

We have measured fringes in only three out of the nine possible detection patterns for $\ket{\psi_4}$ - after the beamsplitter operation, the signal photons can be detected in three states, $\ket{20}$, $\ket{11}$, or $\ket{02}$, which are multiplied by the same three possibilities for the idler photons - but it is in principle possible to monitor all output states simultaneously. This would clearly allow better sensitivity since all the output states provide information about the phase. As described in \cite{nagata}, when calculating the sensitivities from individual fringes, the visibility and the intrinsic detection efficiency of the fringe should be taken into account, as well as its gradient. For the fringes in Fig.~\ref{fringes3}(a), (b), and (c), we find values for $\Delta L$ of 1.18, 1.82, and 1.75 respectively, relative to the SQL. No individual fringe beats the SQL, due to the non-unit visibilities, and because the state is divided between more potential detection patterns than in the single colour case. However, adding the Fisher Information \cite{gioscience} from separate fringes suggests they would allow an improvement if they were monitored simultaneously. Since of the six outputs we did not measure we expect three to be of the same form as (b) and three the same form as (c), we can estimate that monitoring all nine outputs would give a minimum uncertainty 0.72 of the SQL. This would be a significant improvement over the SQL, though it is still above the theoretical value from Eq.~\ref{CR} for $\ket{\psi_4}$ of 0.61 due to the non-unit visibilities, and above the Heisenberg limit for four photons of 0.5.

A general wavefunction for m photons of two colours can be written as
\begin{equation}
\ket{\psi_m}=\sum_{r=0}^{m/2}\frac{ (-1)^r  e^{2ir\theta_p}}{\sqrt{\frac{m}{2} +1}}\ket{m/2 - r, r}_s\ket{m/2 - r, r}_i.
\end{equation}
For example m=6 leads to
\begin{equation}
\ket{\psi_6}=\frac{1}{2}\left(\begin{array}{cc}\ket{30}_s\ket{30}_i-e^{2i\theta_p}\ket{21}_s\ket{21}_i\\ +e^{4i\theta_p}\ket{12}_s\ket{12}_i-e^{6i\theta_p}\ket{03}_s\ket{03}_i \end{array}\right),
\end{equation}
as shown in Appendix C. Figure~\ref{fringes4} shows an example of $\ket{\psi_6}$ with the detection pattern $\ket{21}_s\ket{21}_i$, which is expected to contain oscillation at both two and six times the classical frequency. Multiphoton interference again plays a part - if separate photon pairs were distinguishable, only two-fold oscillation would remain. The data is in good agreement with theory, though there is a high level of background noise from higher-order emission due to the high pump power used and the low count rate. It is clear that the data taken does not achieve the $82\%$ visibility required to exceed the SQL, however improvements in the level and balancing of collection and detection efficiencies would bring us towards this goal.
\begin{figure}[t]
\vspace{-0.4cm}
\includegraphics[width=\columnwidth]{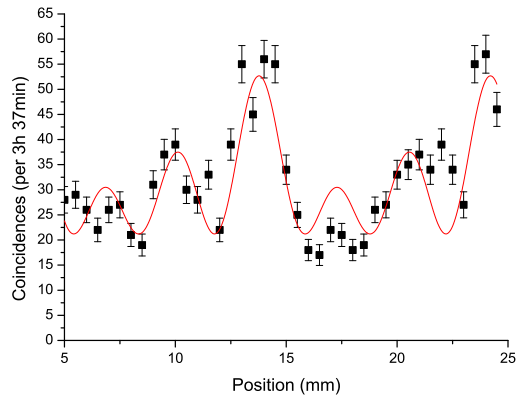}
\vspace{-0.75cm}
\caption{Six fold coincidences in the $\ket{21}_s\ket{21}_i$ output with a fit line based on theory.}
\label{fringes4}
\end{figure}

$\ket{\psi_m}$ is an equally weighted superposition of all distributions of $m/2$ signal-idler pairs between the two paths, with each path necessarily containing equal numbers of signal and idler photons. This results in simple expressions for Eq. 2 and 3 (see Appendix C), leading to
a minimum uncertainty in $L$ given by
\begin{equation}
\Delta L\geq \left(\frac{3c^2\hbar}{\omega_p E (m+4)}\right)^{1/2}.
\end{equation}
Hence for a general two colour entangled state the shot noise will scale below the SQL by a factor $\sqrt{(m+4)/3}$, potentially achieving significant improvements at large $m$. With current technology, detecting large numbers of coincident photons is impractical due to the relatively low efficiency of single photon detectors, and the improved sensitivity of the state drops rapidly with any form of loss including imperfect detectors. Although an analysis of loss tolerance is beyond the scope of this work, the similarity of these states to Holland-Burnett states suggests that they will fare better than NOON states. Also note that the states are generated spontaneously by the source, so that it is not possible to obtain a high sensitivity $m$ photon entangled state on demand. Rather we expect that in a low loss setting, increasing the pair generation rate of the source will improve sensitivity as more high-$m$ states occur, unlike the current situation where a higher generation rate tends to add noise to the results.

In conclusion we have shown that two colour entangled states could be a useful resource for quantum metrology and demonstrated these effects up to six photon states in a novel ultra-stable interferometer. We have seen that single photon interference \textit{and} multiphoton interference combine to increase the visibility of fringes at the four and six photon level, and ultimately lead to improved sensitivity. When lumped losses are reduced giving access to entangled states consisting of larger numbers of photons, the simplicity of this path-entangled pair photon source and its improved sensitivity to path-length make it a promising approach to future quantum metrology and enhanced sensing.

We acknowledge support from UK EPSRC, EU grants 248905 Q-Essence and 600838 QWAD, ERC grant 247462 QUOWSS, and the Australian Research Council Centre of Excellence (CUDOS, CE110001018) and DECRA (DE130101148) schemes.

\section{Appendix A: Experimental Methods}

\begin{figure}[h]
\begin{centering}
\includegraphics[width=\columnwidth]{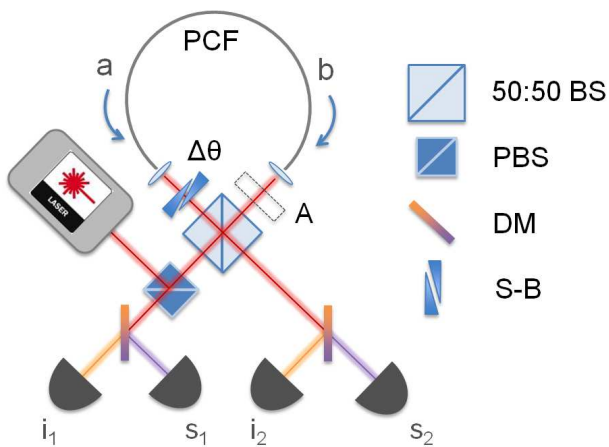}
\caption{Experimental setup. BS: beamsplitter; DM: dichroic mirror; PBS: polarizing beamplitter; S-B: Soleil-Babinet compensator.}
\label{setup}
\end{centering}
\end{figure}

The birefringent PCF source used in this experiment is similar to that described in \cite{halder, clark}, using a cross-polar phase matching scheme such that pump light polarized on the slow axis of the fibre produces signal and idler photons polarized on the fast axis through four wave mixing (FWM). The phase matching curve has a turning point in the signal wavelength, at which the signal spectrum is not correlated with that of the pump or idler. Hence when the fibre is pumped with picosecond pulses at 720~nm (produced from a \emph{Spectra-Physics} Tsunami Ti:Sapphire laser) the joint spectrum of the signal and idler at 625~nm and 860~nm should be free of correlation, resulting in the photons being detected in a pure quantum state without the need for tight filtering and the associated loss.

In order to see classical interference in the Sagnac interferometer, a half wave plate (HWP) at $45^{\circ}$ was inserted in the space labelled A in Fig.~\ref{setup}. Hence the counter-clockwise propagating pump light was rotated from vertical to horizontal polarization before it reached the variable birefringence, accumulating a phase $\theta_{p H}$, while the clockwise propagating light was still vertically polarized and experienced a phase $\theta_{p V}$. When the two paths crossed at the beamsplitter (BS) again, they had both been rotated to horizontal, and interfered with a relative phase given by $\theta_p=\theta_{p H}-\theta_{p V}$, which could be varied with the position of the Soleil-Babinet compensator (S-B). The BS used was an ultra broadband (600-900~nm) cube from \emph{Laser 2000} \cite{laser2000} with a close to 50:50 splitting ratio for all the wavelengths used.

For the multiphoton interference measurements the HWP was removed from the loop, as the polarization rotation was effectively performed by the cross-polar phasematching, taking vertically polarized pump light to horizontally polarized signal and idler. Counter clockwise propagating pump light produced correlated signal and idler photons in the PCF which then reached the S-B and accumulated a total phase $\theta_{s H}+\theta_{i H}$. Clockwise propagating pump experienced a phase $\theta_{p V}$, which then was transferred to signal-idler pairs produced in the PCF with a factor of 2 due to the quadratic dependence of FWM on the pump field, $2\theta_{p V}$. This leaves a total phase between pairs produced in separate paths $\theta_{s H}+\theta_{i H}-2\theta_{p V}\approx2\theta_p$, where we have assumed that $\theta_{s H}+\theta_{i H}\approx2\theta_{p H}$, which is true unless the dispersion in the S-B is extreme. So the two photon curves will have half the fringe spacing of classical interference with the pump laser.

Initially, the visibility of the two photon interference was limited by chromatic dispersion in the S-B, which caused counter-clockwise signal and idler pulses to walk-off from the accompanying pump pulse, and so arrive at the BS at a slightly different time to the signal and idler from the other direction, which did not experience this effect. In order to re-balance these arrival times, a second identical S-B was inserted into position A. This was kept at a fixed position while the phase was varied with the first S-B.

\section{Apendix B: Non-Unitary Beamsplitter Behaviour}

\begin{figure}[h]
\begin{centering}
\vspace{-0.4cm}
\includegraphics[width=\columnwidth]{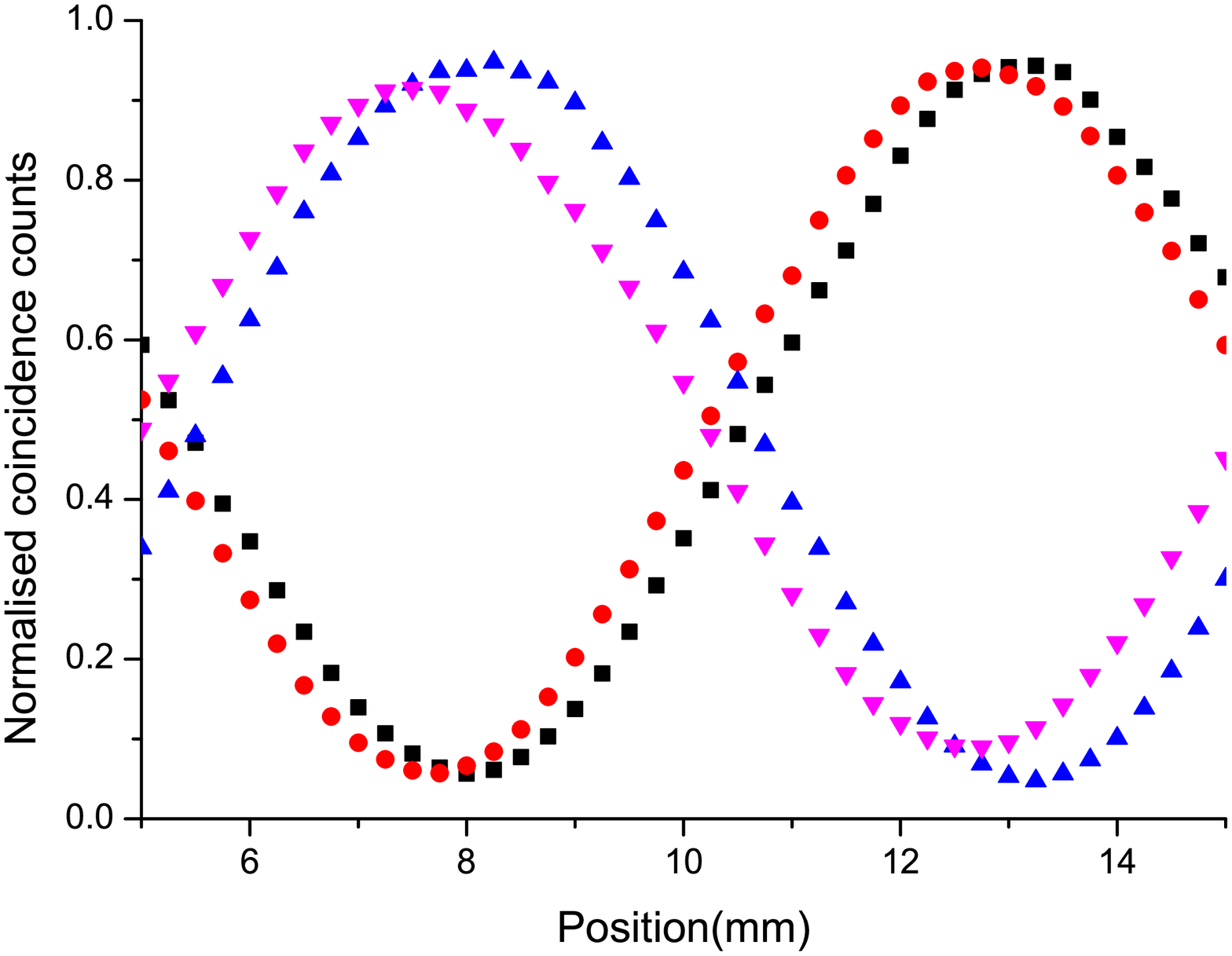}
\vspace{-0.5cm}
\caption{Coincidences between signal and idler (s, i) at output 1 or 2. Black: s1$\wedge$i1; red: s2$\wedge$i2; blue: s1$\wedge$i2; purple: s2$\wedge$i1.}
\label{fringes2}
\end{centering}
\end{figure}
Closer inspection of the two photon coincidence curves (Fig.~\ref{fringes2}) reveals slight phase offsets between fringes, for instance between the counts for a signal and idler photon both exiting output 1 compared to both exiting output 2, which ideally would be exactly in phase. This can be explained by modelling the beamsplitter operation for an individual photon as a Hadamard matrix
\begin{equation}
\left(\begin{array}{lr} 1 & 1 \\ 1 & -e^{i\delta}\end{array}\right)
\end{equation}
which is only unitary if $\delta=0$. From the observed curves, at the signal wavelength $\delta_s\approx$ 0.26~rad and at the idler wavelength $\delta_i\approx$ -0.04~rad. This departure from unitary behaviour is possible if there is sufficient loss $R$ in the operation \cite{bonneau}:
\begin{equation}
sin \left(\frac{\delta}{2}\right)\leq\frac{R}{1-R}.
\end{equation}
Hence a value of 0.26~rad requires a loss $R\geq 11\%$. The loss at the beamsplitter at 635~nm measured with a bright laser is 12.4$\%$. The fact that the effect is much larger at the signal wavelength than the idler suggests it is a wavelength dependent property of the broadband beamsplitter, and that the problem could be avoided by using a smaller wavelength separation between signal and idler. The values of $\delta_{s,i}$ can significantly affect the form of the four and six photon fringes, so the measured values from the two photon data were incorporated into the theory lines.

\section{Appendix C: Length Sensitivity for Higher Photon Numbers}
\begin{figure}[h]
\vspace{-0.4cm}
\includegraphics[width=\columnwidth]{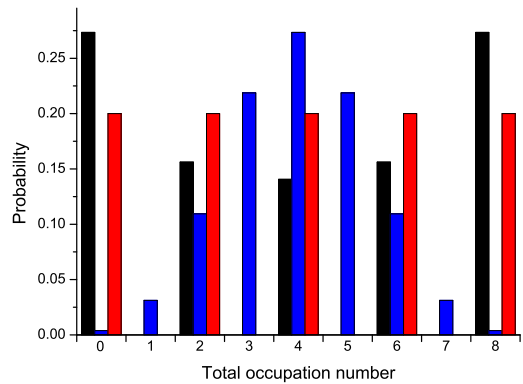}
\vspace{-0.5cm}
\caption{Theoretical probability distributions for total photon number in one arm of the interferometer for three 8 photon states. Black: Holland-Burnett states, produced by a degenerate pair photon source in the interferometer, or Hong-Ou-Mandel interference between indistinguishable photons. Red: When a non-degenerate pair-photon source is used in the interferometer, as in our experiment, the distribution becomes flat. Blue: Uncorrelated photons as in a classical experiment will be distributed binomially. Note that the correlated states only contain even terms as the photons are bunched into pairs.}
\label{probs}
\end{figure}
For a single FWM source pumped by a bright coherent laser, the nonlinear Hamiltonian can be written as
\begin{equation}
g\hat{a}_s^{\dagger}\hat{a}_i^{\dagger}+g\hat{a}_s\hat{a}_i
\end{equation}
where the first term can spontaneously generate a signal-idler pair, and the second term represents the reverse process, which removes pairs. The constant g is a function of the pump power and the nonlinearity of the medium. When the generation rate is low, the reverse process is unlikely to occur and this term can be neglected, so that a propagator can be written as
\begin{equation}
e^{\alpha \hat{a}_s^{\dagger}\hat{a}_i^{\dagger}}=1+\alpha\hat{a}_s^{\dagger}\hat{a}_i^{\dagger}+\frac{\alpha^2}{2!}\hat{a}_s^{\dagger 2}\hat{a}_i^{\dagger 2}+\frac{\alpha^3}{3!}\hat{a}_s^{\dagger 3}\hat{a}_i^{\dagger 3}+...
\end{equation}
where the constants relating to the generation rate have been grouped into $\alpha=-i\hbar gt$ for brevity, with $t$ the interaction time. Applying this propagator to an initial vacuum state for the signal and idler modes, the output state is
\begin{equation}
\ket{vac}+\alpha\ket{1}_s\ket{1}_i+\alpha^2\ket{2}_s\ket{2}_i+\alpha^3\ket{3}_s\ket{3}_i+...
\end{equation}
so that, with identical sources in each arm of the interferometer, any integer combination of $p$ and $q$ pairs produced in each arm can occur with amplitude $\alpha^{p+q}$. For $m/2$ signal and $m/2$ idler photons detected, they are equally likely to have come from any distribution of $m/2$ pairs between the two arms, so the wavefunction for $m$ photons can be written as:
\begin{equation}
\ket{\psi_m}=\sum_{r=0}^{m/2}\frac{ (-1)^r  e^{2ir\theta_p}}{\sqrt{\frac{m}{2} +1}}\ket{m/2 - r, r}_s\ket{m/2 - r, r}_i.
\end{equation}
In this respect, these states differ from Holland-Burnett states, which are weighted slightly towards the wings of the distribution, as in Fig.~\ref{probs}. This results in simple expressions for the operator $\hat{H}_L$ and $\Delta\hat{H}_L$, with
\begin{equation}
H_L = \frac{\omega_s \hat{n}_s + \omega_i \hat{n}_i}{c} = \frac{(\omega_s + \omega_i)\hat{n}_s}{c} = \frac{2\omega_p \hat{n}_s}{c}
\end{equation}
and
\begin{equation}
\Delta H_L = \frac{2\omega_p}{c}\Delta\hat{n}_s
\end{equation}
The uncertainty in the signal photon number is given by
\begin{equation}
\Delta\hat{n}_s^2 = \sum_{r=0}^{m/2} P(n_s=r)r^2 - \left\langle\hat{n}_s\right\rangle^2 = \frac{1}{m/2 + 1}\sum_{r=0}^{m/2}r^2 - (m/4)^2.
\end{equation}
Using $\sum_{i=0}^n i^2 = n(n+1)(2n+1)/6$ we have
\begin{equation}
\Delta H_L =  \frac{\omega_p }{c}\sqrt{\frac{m^2+4m}{12}},
\end{equation}
resulting in a minimum uncertainty in $L$ given by
\begin{equation}
\Delta L\geq \left(\frac{3c^2\hbar}{\omega_p E (m+4)}\right)^{1/2}.
\end{equation}

Importantly for the higher order states, the photon statistics depend on whether the signal photons from separate pairs are distinguishable from each other and whether the idler photons are distinguishable from each other. For instance strong spectral correlations between the signal and idler of a pair can introduce distinguishability. In the extreme case where each photon is completely distinguishable from every other photon, we would expect the output of a single FWM source to produce pairs with Poisson likelihood, such that the generation of one signal-idler pair does not affect the chance of generating another. The flat distribution in Fig.~\ref{probs} then becomes a binomial distribution of pairs - ie. it is still only even terms which are occupied, but there would now be a weighting towards the centre of the distribution. $\Delta\hat{n}_s$ becomes $\sqrt{m/8}$ and $\Delta L$ can only scale a fixed factor of $\sqrt{2}$ below the SQL. Hence no further improvement in resolution is seen for higher photon number states, beyond the initial factor of $\sqrt{2}$ for the two photon state.

\section{Appendix D: Output States and Theoretical Fringes}

For the theoretical fit lines used in the results, we began by deriving the output states from the $\ket{\psi_m}$ states, representing the final beamsplitter operation as a Hadamard operation on the field operators such that $\hat{a}_x^{\dagger}\rightarrow\frac{1}{\sqrt{2}}(\hat{a}_x^{\dagger}+\hat{b}_x^{\dagger})$ and $\hat{b}_x^{\dagger}\rightarrow\frac{1}{\sqrt{2}}(\hat{a}_x^{\dagger}-\hat{b}_x^{\dagger})$, with $\hat{b}_x^{\dagger}$ a creation operator for the sample arm of the interferometer, and $x=s,i$ for the signal and idler modes. These were then modified to account for the non-unitary beamsplitter, with $\hat{b}_x^{\dagger}\rightarrow\frac{1}{\sqrt{2}}(\hat{a}_x^{\dagger}-e^{i\delta_x}\hat{b}_x^{\dagger})$, and the probabilities were used to give the form of the fit lines. A flat background was included as a free parameter.

For $m=2, 4, 6$, we can write the unmodified output states in the photon number basis after the final beamsplitter, omitting global phases, as
\begin{widetext}
\begin{equation}
\ket{\psi_2}_{out} = \frac{cos~\theta}{\sqrt{2}}\ket{1,0}_s\ket{1,0}_i+\frac{sin~\theta}{\sqrt{2}}\ket{1,0}_s\ket{0,1}_i+\frac{sin~\theta}{\sqrt{2}}\ket{0,1}_s\ket{1,0}_i+\frac{cos~\theta}{\sqrt{2}}\ket{0,1}_s\ket{0,1}_i,
\end{equation} 

\begin{equation}
\begin{array}{rl}
\ket{\psi_4}_{out} =& \frac{cos(2\theta)+1}{2\sqrt{3}}\ket{2,0}_s\ket{2,0}_i+\frac{sin~2\theta}{\sqrt{6}}\ket{2,0}_s\ket{1,1}_i +\frac{cos(2\theta)-1}{2\sqrt{3}}\ket{2,0}_s\ket{0,2}_i\\
&+\frac{sin~2\theta}{\sqrt{6}}\ket{1,1}_s\ket{2,0}_i+\frac{cos~2\theta}{\sqrt{3}}\ket{1,1}_s\ket{1,1}_i+\frac{sin~2\theta}{\sqrt{6}}\ket{1,1}_s\ket{0,2}_i\\ &+\frac{cos(2\theta)-1}{2\sqrt{3}}\ket{0,2}_s\ket{2,0}_i+\frac{sin~2\theta}{\sqrt{6}}\ket{0,2}_s\ket{1,1}_i +\frac{cos(2\theta)+1}{2\sqrt{3}}\ket{0,2}_s\ket{0,2}_i,
\end{array}
\end{equation} 

\begin{equation}
\begin{array}{rl}
\ket{\psi_6}_{out} =& \frac{cos(3\theta)+3cos(\theta)}{8}\ket{3,0}_s\ket{3,0}_i+\frac{\sqrt{3}sin(3\theta)+\sqrt{3}sin(\theta)}{8}\ket{3,0}_s\ket{2,1}_i+ \frac{\sqrt{3}cos(3\theta)-\sqrt{3}cos(\theta)}{8}\ket{3,0}_s\ket{1,2}_i\\
&+\frac{sin(3\theta)-3sin(\theta)}{8}\ket{3,0}_s\ket{0,3}_i+\frac{\sqrt{3}sin(3\theta)+\sqrt{3}sin(\theta)}{8}\ket{2,1}_s\ket{3,0}_i+
\frac{3cos(3\theta)+cos(\theta)}{8}\ket{2,1}_s\ket{2,1}_i\\
&+\frac{3sin(3\theta)-sin(\theta)}{8}\ket{2,1}_s\ket{1,2}_i +\frac{\sqrt{3}cos(3\theta)-\sqrt{3}cos(\theta)}{8}\ket{2,1}_s\ket{0,3}_i+
\frac{\sqrt{3}cos(3\theta)-\sqrt{3}cos(\theta)}{8}\ket{1,2}_s\ket{3,0}_i\\
&+\frac{3sin(3\theta)-sin(\theta)}{8}\ket{1,2}_s\ket{2,1}_i +\frac{3cos(3\theta)+cos(\theta)}{8}\ket{1,2}_s\ket{1,2}_i
+\frac{\sqrt{3}sin(3\theta)+\sqrt{3}sin(\theta)}{8}\ket{1,2}_s\ket{0,3}_i\\
&+ \frac{sin(3\theta)-3sin(\theta)}{8}\ket{0,3}_s\ket{3,0}_i+\frac{\sqrt{3}cos(3\theta)-\sqrt{3}cos(\theta)}{8}\ket{0,3}_s\ket{2,1}_i+
\frac{\sqrt{3}sin(3\theta)+\sqrt{3}sin(\theta)}{8}\ket{0,3}_s\ket{1,2}_i\\
&+ \frac{cos(3\theta)+3cos(\theta)}{8}\ket{0,3}_s\ket{0,3}_i.
\end{array}
\end{equation} 
\end{widetext}

From $\ket{\psi_2}_{out}$ we can see that the output state will oscillate between being in one of two `bunched' states where signal and idler are in the same output with probability $cos^2\theta$, and two `anti-bunched states' with probability $sin^2\theta$. When $\delta_{s,i}\neq 0$, these fringes are shifted such that
\begin{equation}
\begin{array}{l}
P[(1,0)_s(1,0)_i]=1/2~cos^2\theta\\
P[(1,0)_s(0,1)_i]=1/2~sin^2(\theta-\delta_i/2)\\
P[(0,1)_s(1,0)_i]=1/2~sin^2(\theta-\delta_s/2)\\
P[(0,1)_s(0,1)_i]=1/2~cos^2(\theta-\delta_s/2-\delta_i/2)
\end{array}
\end{equation}
as can be seen experimentally in Fig.~\ref{fringes2}. Note that these probabilities are no longer correctly normalised, due to the non-unitary nature of the beamsplitter operation.

From $\ket{\psi_4}_{out}$ we can see that the probability of seeing one signal and one idler at each output ($\ket{1,1}_s\ket{1,1}_i$) varies as $1/3~cos^22\theta$. This is modified by $\delta_{s,i}$ such that
\begin{equation}
P[(1,1)_s(1,1)_i]\propto\left[\begin{array}{c} cos\left(2\theta-\frac{\delta_s+\delta_i}{2}\right)\\+sin\left(\frac{\delta_s}{2}\right)sin\left(\frac{\delta_i}{2}\right)\end{array}\right]^2.
\end{equation}
Also from $\ket{\psi_4}_{out}$, we can see that there are four detection patterns where the probability varies as $1/6~sin^22\theta$, all involving three of the photons at one output and the remaining photon at the other output. As an example we have measured the output $\ket{2,0}_s\ket{1,1}_i$. The effect of $\delta_{i}$ changes the corresponding probability to
\begin{equation}
P[(2,0)_s(1,1)_i]\propto[sin(2\theta-\delta_i/2)-sin(\delta_i/2)]^2.
\end{equation}
The remaining four output detection patterns for $\ket{\psi_4}$ occur with probabilities $1/3~[cos(2\theta)+1]^2$ or $1/3~[cos(2\theta)-1]^2$. Like the two photon fringes, the effect of $\delta_{s,i}$ here is only to offset the phase slightly. As an example we measured the output $\ket{2,0}_s\ket{0,2}_i$. We find that the data is a good fit to the modified theoretical curves in all three cases when we use values of $\delta_{s,i}$ taken from the two photon fringes.

From $\ket{\psi_6}$, we measure a fringe for one output, $\ket{2,1}_s\ket{2,1}_i$. This was chosen because it is one of the outputs which contains a large component of oscillation at $6\theta$ as opposed to $2$ or $4 \theta$, so that six-fold super-resolution should be evident. The probability $1/64~[3cos(3\theta)+cos(\theta)]^2$ is modified to
\begin{widetext}
\begin{equation}
P[(2,1)_s(2,1)_i]\propto\left[
\begin{array}{rl}
&3cos\left(3\theta-\frac{\delta_s+\delta_i}{2}\right) +(4-2cos\delta_s-2cos\delta_i+cos(\delta_s+\delta_i))cos\left(\theta-\frac{\delta_s+\delta_i}{2}\right)\\
&+(2sin\delta_s+2sin\delta_i-sin(\delta_s+\delta_i))sin\left(\theta-\frac{\delta_s+\delta_i}{2}\right)
\end{array}
\right]^2
\end{equation}
\end{widetext}
which is also in good agreement with the experimental results.

\end{document}